\documentclass[useAMS,usenatbib]{mn2e}
\usepackage{epsfig} % used for \epsfig in the figure environment
\usepackage{amssymb}          % used for \lesssim
\usepackage[T1]{fontenc}
\usepackage{ae}
\usepackage{times}
\newcommand{\eps}{\epsilon}
\newcommand{\lnL}{\ln\Lambda}
\title[The efficiency of the spiral-in of a black hole to the Galactic
	centre]{The efficiency of the spiral-in of a black hole to the
	Galactic centre}
\author[P.F. Spinnato, M. Fellhauer, S.F. Portegies Zwart]
{Piero F. Spinnato$^a$
\thanks{E-mail: piero@science.uva.nl (PS); mike@astrophysik.uni-kiel.de (MF); spz@science.uva.nl (SPZ)}, %
Michael Fellhauer$^b$ %
and Simon F. Portegies Zwart$^{a,c}$\\
$^a$Section Computational Science, Universiteit van Amsterdam, Kruislaan~403, 1098~SJ~Amsterdam, The Netherlands\\
$^b$Institut f{\"u}r Theoretische Physik und Astrophysik der
Christian-Albrechts-Universit{\"a}t zu Kiel, Leibnizstr.~15, D-24118~Kiel, Germany\\
$^c$Astronomical Institute `Anton Pannekoek', Universiteit van Amsterdam, Kruislaan~403, 1098~SJ~Amsterdam, The Netherlands}
\begin{document}

\date{}
\maketitle
\begin{abstract}

We study the efficiency at which a black hole or dense star cluster
spirals in to the Galactic centre.  This process takes place on a
dynamical friction time scale, which depends on the value of the
Coulomb logarithm ($\lnL$).  We determine the accurate value of
this parameter using the direct $N$-body method, a tree algorithm and a
particle-mesh technique with up to 2~million plus one particles.  The
three different techniques are in excellent agreement. Our result for
the Coulomb logarithm appears to be independent of the number of particles. We
conclude that \mbox{$\lnL = 6.6 \pm 0.6$} for a massive point particle
in the inner few parsec of the Galactic bulge.  For an extended
object, like a dense star cluster, $\lnL$ is smaller, with a
value of the logarithm argument $\Lambda$ inversely proportional to the object size.
\end{abstract}


\section{Introduction}

The very young objects near the Galactic centre, such as the
Quintuplet star cluster~(\citealt{NagWoo90}; \citealt{OkuShi90}), the
Arches cluster~\citep{NagWoo95} and the central star
cluster~(\citealt{TamRie93}; \citealt{KraGen95}) are of considerable
interest. One of the more interesting conundrums is the presence of
stars as young as few Myr~(\citealt{TamRie93}; \citealt{KraGen95})
within a parsec from the Galactic centre~\citep{Ger01}. \mbox{\it In situ} formation is problematic, due to the strong tidal field of the Galaxy, which makes this region inhospitable for star formation. One possible solution is provided by
\citet{Ger01}, who proposes that a $10^6~{\mathrm M}_{\odot}$ star
cluster spirals in to the Galactic centre within a few million years
from a distance $\gtrsim 30$ pc. The infall process is driven by
dynamical friction~\citep{Cha43}. A quantitative analysis of this
model by \citet{McMPor02} confirms Gerhard's results. The main
uncertainty in the efficiency of dynamical friction, and therewith the
time scale for spiral-in, is hidden in a single parameter, called the
Coulomb logarithm $\lnL$. Accurate determination of this
parameter is crucial for understanding this process. Nevertheless, a
precise value of $\lnL$ for the Galactic central region is not
available. In this paper we determine $\lnL$ for the Galactic
centre. We focus on the efficiency of the interaction between an
intermediate mass black hole (BH hereafter) and the stars in the
Galactic central region. In Section~\ref{Sect:discussion} we comment on how this
approach can be applied to star clusters that sink to the Galactic centre.

In the classical study of \citet{Cha43}, dynamical friction is driven by the  drag force experienced by a point mass moving through an infinite medium of homogeneous density. The consequences of finiteness and non-homogeneity have been analysed in various works (see~\citealt{Mao93}; \citealt{MilMer01}). \citet{JusPen03} carried out an analytical study of dynamical friction in inhomogeneous systems, leading to a value of the Coulomb logarithm that depends on the infalling object position. \citet{ColPal98} developed a general theoretical framework for the interaction of a satellite with a primary galaxy, able to describe dynamical friction in finite, inhomogeneous systems. They applied their theory of linear response to orbital decay of satellites onto a spherical galaxy~\citep{Col98} and short-lived encounters with a high-speed secondary~\citep{ColPal98}. They studied evolution of satellites in isothermal spherical haloes with cores~\citep{ColMay99}, extended in~\citet{TafMay03}, treating satellite finite size and mass loss. Still, the original expression of Chandrasekhar is used to model dynamical friction in many astronomical situations (see \citealt{BinTre87},~Section~7.1; \citealt{HasFun02}). The cases we study here are characterised by a point mass, with a very small mass compared to the primary system. Therefore Chandrasekhar's formulation is appropriate in our cases.

Dynamical friction is important for a large variety of astronomical
phenomena, e.g. planet migration~(\citealt{GolTre80};
\citealt{CioBru02}), core collapse in dense star clusters
\citep{PorMak99} or mergers in galaxy
clusters~(\citealt{Mak97}; \citealt{CorMuz97};
\citealt{BosLew99}). The physics of the infall process of a satellite
in the parent galaxy is basically the same as in the case of a BH
spiralling-in to the Galactic centre. The relevant parameters,
however, are quite different in the two cases. For example, an
inspiraling galaxy has finite size, whereas a BH is a point
mass. Dynamical friction also plays an important role in the evolution
of the black hole binary formed after the merging of two
galaxies both hosting a BH at their centre~\citep{MilMer01}. In this case, dynamical friction is
important in the early phase of galaxy merging, when the BHs orbits
converge and become bound.

We determine the value of $\lnL$ for a BH spiralling-in to the
Galactic centre by means of self-consistent $N$-body simulations. This
is by far not an easy task. $N$-body models either lack in the number
of particles (a direct $N$-body code can treat up to about $10^{5}$
particles, compared to $10^{8}$ for the real system) or have to
introduce softening \citep{Aar63} and approximation of
the force calculation (treecode \citep{BarHut86} or particle-mesh
code \citep{HocEas88}). The softening parameter $\eps$ was introduced to limit the strength of the mutual gravitational interaction during close stellar encounters. Without softening, the very high accelerations experienced by the encountering bodies would cause very tiny integration steps, which would result in a substantial freeze of the global system evolution, with consequent dramatic performance degradation. The use of such approximation does not invalidate the numerical results, as long as the simulated system is studied on a time scale shorter than the relaxation time scale (\citealt{BinTre87},~ch.4, see also discussion in Section~\ref{Sect:softening} below). The dynamical friction time scale of the systems we simulate is in all cases shorter than the relaxation time scale, so we can safely use the approximate methods. 

Nevertheless, since close encounters have a relevant effect on dynamical friction, decreasing their strength by means of softening also decreases the strength of dynamical friction, i.e. lowers the value of $\lnL$. The same role of softening is played, in the particle-mesh code, by the grid cell size $l$. 

\begin{figure}
%\centering
    \rotatebox{-90}{\epsfig{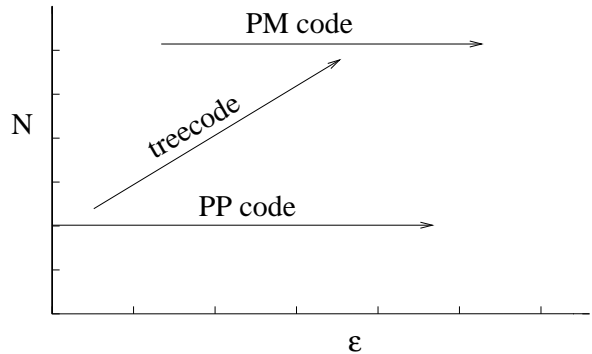}}
\caption{\label{stratefig} 
A sketch of the strategy that we adopt in order to explore the $\eps$-$N$ parameter space.
}
\end{figure}

Our methodological approach for the present work (see Fig.~\ref{stratefig}) consists of comparing the ``exact'' results obtained with the direct method for low particle numbers (up to $80\,000$) with the results of the treecode, which are less accurate and influenced by force softening, to understand how the softening $\eps$ influences the results and how they have to be
scaled according to the value of $\eps$. Then the results of the treecode are compared to the results of the particle-mesh code, to see how softening (tree) 
and grid-resolution $l$ (particle-mesh) can be compared and scaled. Finally,
having the right scaling between the different codes, we will be able
to perform high particle number simulations (up to $4 \cdot 10^{7}$)
with the particle-mesh code to obtain the value of the Coulomb
logarithm for the inner Galactic Bulge.

\section{Methods and model}

\subsection{Direct method}

For our direct $N$-body calculations we used the {\tt kira} integrator
module of the Starlab software environment%
\footnote{See: {\tt http://manybody.org}}
\citep{PorMcM01}. Conceived and written as an independent
alternative to Aarseth's NBODY4 and NBODY5 \citep{Aar85, Aar99}, the
workhorses of collisional $N$-body calculations for the past 25 years,
{\tt kira} is a high-order predictor-corrector scheme designed for
simulations of collisional stellar systems. This integrator
incorporates a Hermite integration scheme \citep{MakAar92} and
a block time step scheduler \citep{McM86} that allows homogeneous
treatment of all objects in the system.

While {\tt kira} is designed to operate efficiently on general-purpose
computers, it achieves by far its greatest speed when combined with
GRAPE-6 special purpose hardware \citep{MakFuk02}%
\footnote{See: {\tt http://www.astrogrape.org}}.
For the work presented here we performed simulations with the GRAPE-6 system at the University of Tokyo with up to $80\,000$ particles.

\subsection{Treecode}

The hierarchical treecode~\citep{BarHut86} is widely used for the
simulation of collisionless systems. The force on a given particle $i$
is computed by considering particle groups of larger and larger size
as their distance from $i$ increases. Force contributions from such
groups are evaluated by using truncated multipole expansions. The
grouping is based on a hierarchical tree data structure. Such
hierarchical tree is realised by inserting the particles one by one
into the initially empty simulation cube. Each time two particles are
into the same cube, this is divided into eight 'child' cubes, whose
linear size is one half of its parent's. This procedure is repeated
until each particle finds itself into a different cube. Hierarchically
connecting such cubic cells according to their parental relation
leads to the hierarchical tree data structure. When force on particle
$i$ is computed, the tree is traversed looking for cells that satisfy
an appropriate acceptance criterion; if the cell is not accepted, then
its children are checked. See~\citet{SalWar94} for a detailed
overview on acceptance criteria. By applying this procedure
recursively starting from the tree root, i.e. the cell containing the
whole system, all the cells satisfying the acceptance criterion are
found.

Our treecode simulations were initially performed with both a code
written by Jun Makino~\citep{Mak91}, and with GADGET~\citep{SprYos01}. In
GADGET each particle is assigned an individual time-step, and at each
iteration only those particles having an update time below a certain
time are selected for force evaluation. This criterion was originally
introduced in the direct $N$-body code ({\it cf.}~\citealt{Aar99}).
This code is parallelized using MPI~\citep{MPI2}. In the parallel version, the geometrical domain is partitioned, and each processor hosts the particles located
in the domain partition assigned to it. The computation of forces on
the selected $i$-particles is performed by scattering the particle
data to remote processors. Then partial forces from the particles
hosted by the remote processors are computed locally. Finally,
calculated forces are received back by the $i$-particle host, and
added up resulting in the total force on the $i$-particles.

\subsection{Particle-mesh code}
\label{sec:superbox}

\begin{figure}
%\centering
  \epsfig{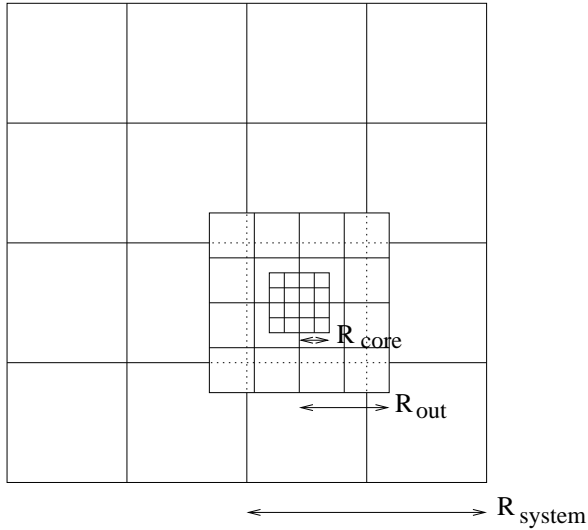}
\caption{\label{SBgridfig} The different grids of {\sc Superbox} for a number of cells per dimension $n = 4$. The finest and intermediate grids are focussed on the object of interest. Each grid is surrounded by a layer of two halo cells. Such haloes are not shown here.}
\end{figure}

To perform calculations using several millions of particles we use a
particle-mesh (PM) code named {\sc Superbox} \citep{FelKro00}.  With
the particle-mesh technique densities are derived on Cartesian grids.
Using a fast Fourier transform algorithm these densities are
converted into a grid-based potential.  Forces acting on the particles
are calculated using these grid-based potentials, making the code
nearly collisionless. The current implementation completely neglects
two-body relaxation causing it to retain only a small amount of
grid-based relaxation \citep{FelKro00}.

The adopted implementation incorporates some differences to standard
PM-codes.  State of the art PM-codes are using a cloud-in-cell (CIC)
scheme to assign the masses of the particles to the grid-cells; which
means that the mass of a particle $i$ is split up into neighbouring
cells according to its distance to the centre of the cell.  Forces are
then calculated by adding up the same fractions of the forces from all
cells to particle $i$.  In contrast our code uses the ``old-fashioned''
nearest-grid-point scheme, where the total mass of the particle
is assigned to the grid-cell the particle is located in.  Forces
acting on the particle are then calculated only from the forces acting
on this particular cell.  To achieve similar precision as CIC, we use
space derivatives up to the second order.

To achieve high resolution at the places of interest, we incorporate
for every simulated object (e.g.\ each galaxy and/or star cluster or
disc, bulge and halo) two levels of sub-grids which stay focused on
the objects of interest while they are moving through the simulated
area (see Fig.~\ref{SBgridfig}). This provides higher resolution only
where it is necessary. Our PM-code is parallelized using MPI.

\subsection{The theory of the Coulomb logarithm}

Dynamical friction affects a mass moving in a background sea of
lower mass objects. A practical expression for the strength of the
drag force on a point particle with mass $M_{BH}$ is
(\citealt{BinTre87},~p.~424):
\begin{equation}
    \frac{d \mathbf{v}_{BH}}{dt} = - 4 \pi G^2 \lnL \rho M_{BH}
    \frac{\mathbf{v}_{BH}}{v_{BH}^3} \left[   
    \mathrm{erf}(X) - \frac{2X}{\sqrt{\pi}}\, e^{-X^2} \right].
\end{equation}
Here $X = v_{BH}/(\sqrt{2} \sigma)$, $\sigma$ being the Maxwellian
velocity dispersion, and $\rho$ is the background stellar density.

The classical value of $\Lambda$ is (\citealt{BinTre87},~p.~423)
\begin{equation}
    \Lambda = \frac{b_{max} v_{typ}^2}{G (M_{BH} + m)}.
\label{loleq} \end{equation}
Here $b_{max}$ is the largest possible impact parameter for an
encounter between the massive point particle and a member of the
background population, $v_{typ}$ is the typical speed of the objects in
the background population, and $m$ is the mass of each of the background stars. Equation~(\ref{loleq}) can then be
generalised to 
\begin{equation}
    \Lambda = \frac{b_{max}}{b_{min}}\, .
\label{lolgeneq} \end{equation}
Here $b_{min}$ is the distance below which an encountering particle is
captured, instead of being scattered by the massive object. It is
somewhat smaller than the $90\degr$ turn-around distance.  With the
direct $N$-body technique, $\Lambda$ can be measured precisely. However,
with approximate $N$-body methods, such as the treecode or the PM code,
we have to take care of the interference of the softening length/cell
size with $b_{min}$, as discussed in Section~\ref{softsec}.

\citet{McMPor02} obtained an analytic expression for the distance $r(t)$ of the
BH to the Galactic centre, with the assumptions that the BH's orbits are nearly
circular, and the mass profile of the Galaxy is given by a power law:
\begin{equation}
        M(R) = AR^\alpha \, .
\label{mprofeq} \end{equation}
They obtained:
\begin{equation}
  r(t) = R_0\left[1 - \frac{\alpha(\alpha+3)}{\alpha+1}
                \sqrt{\frac{G}{A R_0^{\alpha+3}}}\, \chi\, M_{BH}\, \ln
                \Lambda\, t \right]^{\frac{2}{3+\alpha}} \ ,
\label{Rvsteq} \end{equation}
where
\[
  \chi = \mathrm{erf}(X) - \frac{2X}{\sqrt{\pi}}\, e^{-X^2}\ \ \mathrm{and}\ \
   X = \frac{v_{BH}}{\sqrt{2}\, \sigma} \ ,
\]
$\sigma$ being the velocity dispersion. In~\citet{McMPor02} the
value of $X$ in the Galactic centre is also computed, resulting in \mbox{$X = 
\sqrt{2 - \alpha}$}. Finally, we take $R_0$ equal to the half-mass radius $R_{hm}$ (see Section~\ref{incondsec}). The best fit of equation~(\ref{Rvsteq}) on the simulation data gives the value of $\lnL$ for that simulation. Such values, for all simulation performed, are reported in the last column of Tables~\ref{allrunPPtab},~\ref{allrunTREEtab} and~\ref{allrunPMtab}.

\subsection{The role of softening in the determination of the Coulomb logarithm} \label{softsec}
As already mentioned in the introduction, softening was introduced in numerical stellar dynamics to limit the strength of mutual forces during close stellar encounters, mainly for computational performance purposes. It consists in a modification of the Newton law for the gravity exerted by a particle $j$ on a particle $i$, as follows:
\begin{equation}
  \mathbf{a}_i = G \frac{m_j}{(r_{ij}^2 + \eps^2)^{(3/2)}} \mathbf{r}_{ij}
\label{softeq} \end{equation}
where $\mathbf{r}_{ij} = \mathbf{r}_{j} - \mathbf{r}_{i}$, and $\eps$ is the softening parameter. As \mbox{$\mathbf{r}_{ij} \to 0$}, the presence of $\eps$ causes the force to change from inverse square to elastic, with constant $G m_i m_j/\eps^3$. In this way the strength of the mutual force between encountering particles is no more unbound.

Softening was first used by \citet{Aar63} in a particle-particle (PP) context. Accuracy requirements soon led to a more precise treatment of close encounters and binaries by means of an analytic approach (\citealt{KusSti65}; \citealt{Aar72}; \citealt{MikAar90}). The softened force in equation~(\ref{softeq}) remains used in the treecode, where high accuracy in close encounters treatment is not mandatory. Here we will use the softening both in the treecode simulations, where it is necessary, and in the PP code simulations, where it is used to compare the results of the two codes, in order to study the relation between $\eps$ and $\lnL$.

For the PM code, as described in Section~\ref{sec:superbox}, force is not computed by using the Newton force, or the softened force in equation~(\ref{softeq}). Instead, the fact that the gravitational potential on each grid point of the mesh is obtained from a density field defined on the same mesh, leads to an accuracy for the force on each particle limited by the cell size of the grid, $l$.

Here, we are concerned with the accuracy of the computation of the encounters experienced by a black hole spiralling-in to the Galactic centre. Since the softening (PP and treecode) and the cell size (PM code) affect this accuracy, we will use $\eps$ and $l$ to quantify the accuracy decrement in our simulations. In Section~\ref{lnLsec} we will study quantitatively the dependence of $\lnL$ on $\eps$ and $l$. 

The reference value for $\eps$ in the work presented here will be \mbox{$\eps_0 = 0.003735$} (units given below in Table~\ref{unitab}). This value, according to \citet{AthFad00}, is of the same order of magnitude as the optimal softening for a Dehnen sphere distribution \citep{Deh93}. This distribution is similar to the power law distribution that we use, at least for what concerns the high central density peak, which is the key physical factor in the determination of the optimal softening. For an 80\,000 particle distribution, $\eps_0$ is about 15 times smaller than the mean inter-particle distance $\ell$ at the initial BH position \mbox{$R_0 \simeq 0.87$} (see Section~\ref{incondsec}). This value for $\eps$ is small enough to avoid spurious effects in the force between a star and its neighbours, but is sufficient to inhibit very close encounters. The expression for $\ell$ can be obtained as:
\begin{equation}
   \ell = n^{-\frac{1}{3}} = \left(\frac{\rho}{m}\right)^{-\frac{1}{3}} =
             \sqrt[3]{\frac{4 \pi\, R^{3-\alpha}}{N A \alpha}}  
\label{elaveq} \end{equation}
where $n$ is the star number density, and
\[
   \rho = \frac{1}{4 \pi R^2} \frac{dM}{dR} = 
          \frac{A \alpha}{4 \pi} R^{\alpha - 3} \ .
\]
We used the expression in equation~(\ref{mprofeq}) for $M$, and the fact that the $N$ stars in the system have the same mass \mbox{$m = 1/N$}.

One of the effects of softening is a damping in the BH infall at very small values of the galactocentric distance, more noticeable as $N$ increases. This can be explained with the fact that the inter-particle distance $\ell$ decreases as the BH approaches the Galactic centre (see equation~\ref{elaveq}). When $\ell$ becomes comparable to $2\eps$, the role of softening in the force equation becomes dominant, since particles begin to `overlap'. With \mbox{$N = 400\,000$}, we get \mbox{$\ell = 2\eps_0$} when $R \simeq 0.064$, which is close
to the value at which the damping arises, as Fig.~\ref{varmfig} below clearly shows. 

\subsection{Initial condition} \label{incondsec}

\begin{figure}
\centering
    \rotatebox{-90}{\epsfig{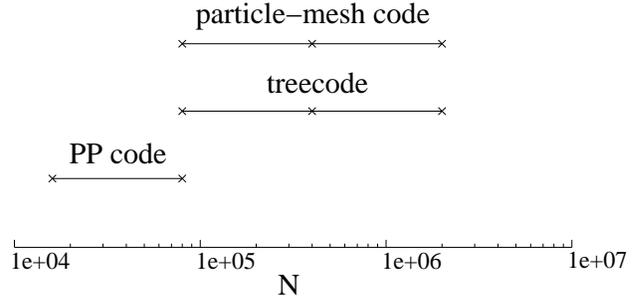}}
\caption{\label{Nrangefig} 
Particle ranges for the simulations performed by each method. Crosses
denote the particle values used.}
\end{figure}

We generate the initial mass distribution according to the power law given by equation~(\ref{mprofeq}), with \mbox{$\alpha = 1.2$}, which reproduces the mass distribution in the centre of the Galaxy, according to~\citet{MezZyl99}. The scale factor is \mbox{$A = 4.25 \cdot 10^6 {\rm M}_\odot$}, corresponding to $0.44$ in the $N$-body standard units~\citep{HegMat85}, which are reported in
Table~\ref{unitab}. We use the standard units hereafter, unless other
units are explicitly reported. The distributions that we generate are truncated at \mbox{$R$ = 1.7 = 13.6\,pc}, with a total mass within this
radius $M_{tot} = 1$. The particles have equal mass $m$. Particles are assigned Maxwellian velocities, then the system is virialised to dynamical
equilibrium. Then, before inserting the black hole (BH) particle, we let the system evolve for a few crossing times. The system reaches a stable configuration, whose mass profile is no more perfectly reproduced by Eq.~\ref{mprofeq}. The best fit for $A$ and $\alpha$ on the mass profile of the stable configuration gives:

\begin{eqnarray}
\label{eq:plcoeffs}
   A      & = &   0.53 \\
   \alpha & = &   0.9 \ . \nonumber
\end{eqnarray}

In fact, the mass profile having these coefficients diverges from the original one as the distance $R$ increases. On the other hand, in the region \mbox{$R < 2$}, where we study the BH infall, the discrepancy between the two mass profiles is small. The relaxed profile values are within 10\% of the initial profile values. Nevertheless, for consistency we will use the values in equation~(\ref{eq:plcoeffs}) for $A$ and $\alpha$ hereafter. This results in values of \mbox{$\lnL \simeq 10$\%} smaller than the ones given by a mass profile with coefficients \mbox{$\alpha = 1.2\ \rmn{and\ } A = 0.44$}.

The BH particle is placed at the half-mass radius $R_{hm} \simeq 0.87$ with a circular orbit velocity, and its mass is $M_{BH} = 0.000528$. The background particles number varies from 16\,000 to 2~million. The low particle number simulations are performed with the PP code, the intermediate and high number simulations with the treecode and the PM code. Fig.~\ref{Nrangefig} shows the particle range for each code. This allows us to span a large range in particle number, so that the influence of granularity in the BH motion towards the Galaxy centre can be studied.

In contrast to the other models, we choose physical units for the PM code simulations. The conversion factors from physical units to
$N$-body units are shown in Table~\ref{unitab},
\begin{table}
\caption{\label{unitab} Conversion table between the $N$-body units used in our work, and physical units. The $N$-body units are such that $G = 1$, $M_{tot} = 1$, and $E_{tot} = -0.25$.}
\begin{eqnarray*}
\label{eq:unitconv}
   G & = & 1 \ {\rm v.u.}^{2}{\rm l.u.}/{\rm m.u.} \nonumber \\
     & \equiv & 4.3007 \cdot 10^{-3} \ {\rm km}^{2}{\rm pc}/{\rm s}^{2}
                {\rm M}_{\odot} \\
     & \equiv & 4.49842 \cdot 10^{-3} \ {\rm pc}^{3}/{\rm Myr}^{2}
                {\rm M}_{\odot} \nonumber \\
   1 \ {\rm l.u.} & = & 8 \ {\rm pc} \\
   1 \ {\rm m.u.} & = & 1.18 \cdot 10^{8} \ {\rm M}_{\odot} \\
   1 \ {\rm v.u.} & = & \sqrt{\frac{G \cdot 1 \ {\rm m.u.}}{1 \ {\rm l.u.}}} 
                         = 251.86 \ {\rm km}/{\rm s} \\
   1 \ {\rm t.u.} & = & \sqrt{\frac{{1 \ {\rm l.u.}}^{3}}{G \cdot 
                        1 \ {\rm m.u.}}} =  0.031 \ {\rm Myr} 
\end{eqnarray*}
\end{table}
where l.u.\ denotes the unit length in $N$-body units, m.u.\ the unit mass, 
v.u.\ the unit velocity and t.u.\ the unit time.

The parameters of the PM calculations are chosen in the following way:
the grid sizes are kept constant at
\begin{eqnarray}
\label{eq:gridsize}
   R_{\rm system} & = & 140.0 \ {\rm pc} \nonumber \\
   R_{\rm out}    & = &   9.6  \ {\rm pc} \\
   R_{\rm core}   & = &   2.4  \ {\rm pc} \nonumber
\end{eqnarray}
and are focussed on the center of mass of the ``bulge'' model, as
sketched in Fig.~\ref{SBgridfig}. To change the resolution we alter
the number of grid cells per dimension from $32$ up to $128$.  With
this choice the cell sizes listed in Table~\ref{tab:gridresol} are
achieved.

\begin{table}
  \caption{Resolutions (i.e. cell sizes) of the different grid levels for
     the different choices of $n$ in the PM code. $n$ denotes the number
     of cells per dimension. The cell sizes of the different grid-levels
     (outer, middle and inner) are given in pc.}
  \label{tab:gridresol}
  \begin{center}
     \begin{tabular}{crrr}
        $n$   & outer   & middle & inner \\ \hline
        $32$  & $10.00$ & $0.69$ & $0.17$ \\
        $64$  & $4.67$  & $0.32$ & $0.08$ \\
	$128$ & $2.26$  & $0.15$ & $0.04$
     \end{tabular}
  \end{center}
\end{table}

To avoid a self-acceleration of the black hole we choose $H_{0,0,0} = 1.0$ of 
the Greens-function, which worsens the energy conservation of the code 
(see~\citealt{FelKro00}) especially in simulations with high particle
numbers per grid cell, but suppresses the forces of particles in the 
same cell on each other and on themselves.

To speed up the simulations, the time step in the PM code simulations
should be as large as possible, but small enough to prevent spurious
results.  Therefore we started with a time step of 1,000~yr and
reduced it to 200 and 50~yr. The results of the 200~yr and 50~yr time
step do not differ from each other, therefore the global time step is
chosen to be 200~yr. Conversely, the time step in the treecode and
direct code simulations is variable and different for each
particle. Time step values are in this case in the range
2--30\,000~yr, with about 90\% of them in the range 100--300~yr.

\begin{table}
  \begin{center}
    \caption{Overview of the PP runs. $N$ is the number of particles, $\eps$ is the softening parameter, $\eps_0 = 0.003735$, $M_{BH}/m$ is the ratio between the BH mass and a particle mass, and $\eps/b_{min}$ the ratio between the softening parameter and the minimal impact parameter.}
    \label{allrunPPtab}
    \begin{tabular}{rrrrr}
        $N$ &  $\eps/\eps_0$ &  $M_{BH}/m$ & $\eps/b_{min}$ & $\lnL$ \\
      \hline \hline
       16K  &      0         &    8.5      &      0         &   3.8 \\
       16K  &      1         &    8.5      &      2.6       &   3.6 \\
      \hline           
       80K  &      0         &   42.3      &      0         &  6.6  \\
       80K  &      0.01      &   42.3      &      0.03      &  6.0  \\
       80K  &      0.1       &   42.3      &      0.3       &  5.3  \\
       80K  &      1         &   42.3      &      2.6       &  4.8  \\
       80K  &      2         &   42.3      &      5.3       &  3.5  \\
       80K  &      8         &   42.3      &     21.2       &  2.8  \\
       80K  &     16         &   42.3      &     42.4       &  1.8  \\
      \hline
    \end{tabular}
  \end{center}
\end{table}
\begin{table}
  \begin{center}
    \caption{Overview of the treecode runs. Meaning of symbols is the same as in Table~\ref{allrunPPtab} above.}
    \label{allrunTREEtab}
    \begin{tabular}{rrrrr}
        $N$ &  $\eps/\eps_0$ &  $M_{BH}/m$ & $\eps/b_{min}$ & $\lnL$  \\
      \hline \hline
       80K  &        1       &   42.3      &    2.6         &  4.7  \\
      400K  &        1       &  211.3      &    2.6         &  5.0  \\
        2M  &        1       & 1056.5      &    2.6         &  4.9  \\
      \hline
       80K  &      0.1       &   42.3      &    0.3         &  5.7  \\
       80K  &        2       &   42.3      &    5.3         &  4.1  \\
       80K  &        8       &   42.3      &   21.2         &  3.0  \\
       80K  &       16       &   42.3      &   42.4         &  2.0  \\
       80K  &       32       &   42.3      &   84.7         &  1.6  \\
      \hline
       80K  &        1       &   84.5      &    1.3         &  5.4  \\
       80K  &        1       &  169.0      &    0.7         &  4.6  \\
      \hline
      400K  &        1       &  422.6      &    1.3         &  4.6  \\
      400K  &        1       &  845.2      &    0.7         &  4.2  \\
      \hline
    \end{tabular}
  \end{center}
\end{table}
\begin{table}
%  \begin{center}
    \caption{ Overview of the PM runs. $N$ is the number of particles, $n$ the number of grid cells per dimension, $m$ the particle mass, $l$ the intermediate grid cell size, $N_c$ the average number of particles per cell, $m_{c}$ the average mass of a cell, $M_{BH}/m_{c}$ the ratio between the BH mass and the cell mass, and finally $l/b_{min}$ the ratio between the cell size and the minimal impact parameter.}
    \label{allrunPMtab}
    \begin{tabular}{rrrrrrrrr}
      $N$ &  $n$ &  $m$ &  $l$ &$N_{c}$& $m_{c}$ & $\frac{M_{BH}}{m_{c}}$ & $\frac{l}{b_{min}}$ & $\lnL$ \\
          &      &[M$_{\odot}$] & [pc] & $[\frac{\#}{\rm cell}]$ & [$\frac{M_{\odot}}{\rm cell}$]
          & & & \\ 
      \hline \hline
      80K &  16 & 1475 & 1.60 &  46.3 & 68287.0 &   0.9 & 114.3 & n/a \\
      \hline
      80K &  32 & 1475 & 0.69 &   3.6 &  5375.4 &  11.6 &  49.3 & 1.9 \\
     400K &     & 295 & 0.69 &  18.2 &  5375.4 &  11.6 &  49.3 & 2.1 \\
       2M &     &   59 & 0.69 &  91.1 &  5375.4 &  11.6 &  49.3 & 2.2 \\
      \hline
      80K &  64 & 1475 & 0.32 &   0.4 &   546.3 & 114.5 &  22.9 & 3.0 \\
     400K &     &  295 & 0.32 &   1.9 &   546.3 & 114.5 &  22.9 & 3.4 \\
       2M &     &   59 & 0.32 &   9.3 &   546.3 & 114.5 &  22.9 & 3.0 \\
      \hline
      80K &  128 & 1475 & 0.15 &  0.04 &    61.9 & 1011  & 10.7  & 2.8 \\
     400K &      &  295 & 0.15 &   0.2 &    61.9 & 1011  & 10.7  & 3.7 \\
       2M &      &   59 & 0.15 &   1.0 &    61.9 & 1011  & 10.7  & 3.8 \\
      \hline
      2M &  256 &   59 & 0.076&   0.1 &     7.4 & 8483  &  5.4  & 4.1\\
      \hline
    \end{tabular}
%  \end{center}
\end{table}

\section{Results}

We will now study the dependence of our results on the number of particles $N$ 
in Section~\ref{Nscalesec}, and compare the various $N$-body methods with 
identical initial realisations in Section~\ref{Sect:comparison}.  After having 
convinced ourselves that the various techniques produce consistent results, we
continue by studying the effect of softening/cell size
(Section~\ref{Sect:softening}) and black hole mass 
(Section~\ref{Sect:mass}) on the value of the Coulomb logarithm in the
inner part of the Galaxy.

Our simulations aimed at several goals. 1) understanding the scaling of the 
system dynamics with respect to the number of particles $N$, and within this 
scaling, how results from different methods compare with each other. 2) How, 
at a fixed value of $N$, the softening parameter influences the dynamics, 
changing the value of $\lnL$. The particle-mesh method does not make use of 
softening. The cell size in the PM code can be seen in this context as a 
softening length. In our framework, it is crucial to understand the relation 
between the PP code and treecode softening parameter and the PM code cell size.
3) We also study how the BH mass influences the infall time. We doubled and 
quadrupled the BH mass, and observed how this affects the value of $\lnL$.  

A resume of all the runs that we performed is reported in Table~\ref{allrunPPtab} for the PP code runs, Table~\ref{allrunTREEtab} for the treecode runs, and finally Table~\ref{allrunPMtab} for the PM code runs. In all of our runs, the system remains in equilibrium during the whole BH infall, with no significant mass loss from stellar escapes, and a mass profile independent of time.

\subsection{Dependence of $\bmath{\lnL}$ on $\bmath{N}$}  \label{Nscalesec}

\begin{figure}
\centering
  \epsfig{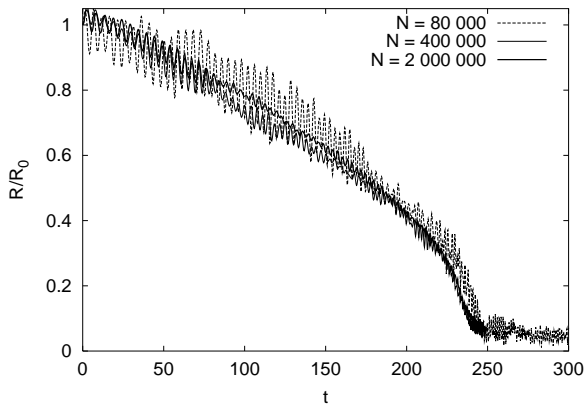}
\caption{\label{Ncompfig} Time evolution of the radial distance of the
black hole to the Galactic centre. The various curves (identified in
the top right corner) present results obtained with the treecode.
the X-axis is presented in $N$-body time units: one $N$-body time unit
corresponds to about 0.031~Myr. The distance of the black hole to the
Galactic centre (Y-axis) is given in terms of its initial distance.
In these simulations is \mbox{$\eps = 0.003735 \simeq 0.03$~pc} and
\mbox{$M_{BH} = 0.000528$}.}
\end{figure}

\begin{figure}
\centering
  \epsfig{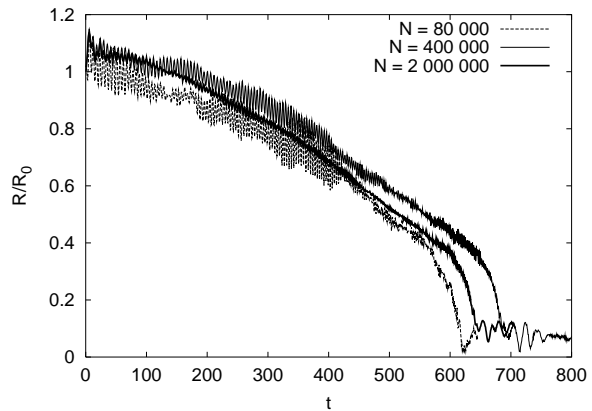}
\caption{\label{sboxNscalfig} Same as Fig.~\ref{Ncompfig} above, but for PM code simulations. The intermediate grid cell size is here \mbox{$l = 0.69$~pc}, and \mbox{$M_{BH} = 0.000528$}.}
\end{figure}

In order to obtain a precise measure of $\lnL$, ideally one
would run a direct $N$-body simulation with $N$ of the order of the
number of stars in the Galactic bulge, which amounts to~\mbox{$\sim
10^8$}. Such high number makes a direct simulation unfeasible, and
imposes the use of approximate methods instead. In order to evaluate
the reliability of the approximate methods, we compared the PP code
runs with the treecode runs. The PP code runs give a reliable picture
of the system dynamics at low particle numbers, i.e. at high
granularity. Using the treecode we can reach a much highernumber of particles, up to two million, which still is two orders of magnitude lower than the
real system. A comparison of the results from the two methods
allows us to estimate the validity of the treecode runs, up to
2~million particles. Then we can compare the treecode runs and the PM runs, 
in order to validate the results from the latter,
which has the capability to simulate systems of about 100~million
stars. In such a way we will eventually be able to study the infall of a BH 
into the Galactic centre in a simulation environment with a realistic value of
$N$. 

In Fig.~\ref{Ncompfig} we show the evolution of the BH distance from
the centre of mass of the system for three treecode simulations. $N$
varies from 80\,000 to 400\,000 and 2~million, with \mbox{$\eps =
\eps_0 = 0.003735$}, corresponding to about 0.03\,pc. In
Fig.~\ref{sboxNscalfig} we present a similar figure from PM code
simulations. Here is \mbox{$N \in \{80\,000, 400\,000, 2\,000\,000\}$}, with 32
cells per dimension, resulting in a cell size of about 0.69\,pc.

Figs~\ref{Ncompfig} and~\ref{sboxNscalfig} show that increasing $N$
results in a much smoother motion of the BH in its infall towards the
centre of the Galaxy. The BH infall rate is not much affected by a change in 
$N$. Accordingly, the value of $\lnL$ for each of the two sets above is 
consistent, as values in Table~\ref{allrunTREEtab} (first three rows) and 
Table~\ref{allrunPMtab} (rows with \mbox{$l = 0.69$}) show. 

\begin{figure}
\centering
  \epsfig{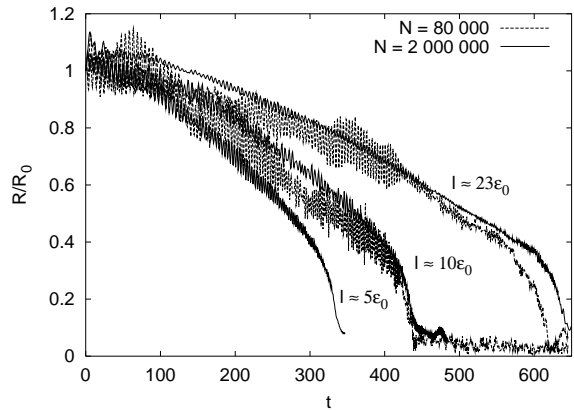}
\caption{\label{rescalfig} Black hole infall at various cell sizes, and 
large difference in $N$. Results here are from PM code simulations. The case 
\mbox{$N = 80\,000$}, \mbox{$l \simeq 5\eps_0$} is not shown for readability 
reasons, since it would overlap with the \mbox{$l \simeq 10\eps_0$} results.}
\end{figure}

In order to study further the extent of the influence of $N$ on the infall rate
of the BH, and hence in $\lnL$, we compare in Fig.~\ref{rescalfig} results from
PM code simulations with increasing grid refinement, and extreme difference in 
$N$. To quantify the grid resolution, we use the  cell length at intermediate 
refinement, which is the cell length pertaining to the physical region where 
the BH evolves for most of its infall. We measure this length in units of 
$\eps_0 = 0.003735$, which makes the comparison with the softening parameter of
the treecode easier. $N$ has no remarkable influence on the infall rate, except
for the case where the cell size is \mbox{$l = 0.15\,\rmn{pc} \simeq 5\eps_0$}.
In this case the simulation with \mbox{$N = 80\,000$} ((data not reported in 
the figure), shows an incorrect BH infall, comparable to the case \mbox{$l = 
0.32 \simeq 10\eps_0$}. This can be explained by the fact that in the low~$l$, 
low~$N$ case, the cells are so small, and the particles so few, that many cells
in the PM grid are empty (see also the $N_c$ column in table~\ref{allrunPMtab},
which gives the average number of particles per cell).
When \mbox{$N_c \ll 1$}, the density field is incorrect, with many grid points 
having a null value, because the corresponding cell is empty. In this 
condition, the gravity field computed by the PM code becomes unreliable, 
affecting the numerical results, as in the simulation with \mbox{$N = 80\,000$}
and \mbox{$l \simeq 5\eps_0$}.

\subsection{Comparison of the codes}\label{Sect:comparison}

\begin{figure}
\centering
  \epsfig{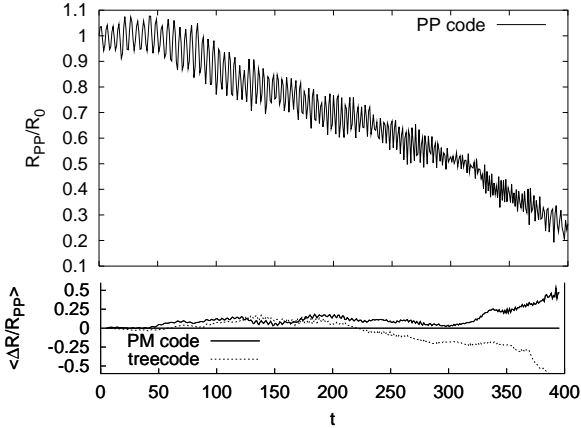}
\caption{\label{obs_fig} Top panel shows a black hole infall simulated by the PP code, with \mbox{$N = 80\,000$}, \mbox{$M_{BH} = 0.000528$} and \mbox{$\eps = 8\eps_0$}. Bottom panel shows a comparison of the PP results with treecode and PM results. Parameter values are in all cases the same, except for the PM cell size, which is \mbox{$l = 10\eps_0$}. Plotted values are averages over 10 time units.}
\end{figure}

In this section we compare the results obtained from the various codes, to check their consistency. The comparison of the PM results with the two other codes results is particularly critical, since the PM code computes forces using a different mathematical approach, i.e.\ a grid based force derivation vs a direct particle-particle computation for the PP code, or particle-multipole computation for the treecode. A consequence of this is a different parameter to tune the accuracy of the simulation, namely the cell size $l$ for the PM code, and the softening length $\eps$ for the other two codes. We will study here how these two parameters influence the black hole infall.

In Fig.~\ref{obs_fig} we show the time evolution of the galactocentric BH distance $R$ simulated by the PP code, accompanied by a plot of the time evolution of \mbox{$\Delta R/R_{PP}$} for treecode and PM simulations, where  \mbox{$\Delta R = (R - R_{PP})$}. The relative difference $\Delta R/R_{PP}$ remains small for a large fraction of the infall, and the final discrepancy is mostly due to the small values of the quantities at that point, which are likely to amplify relative differences. As the following figures also show, the BH infall is predicted with very good consistency among the codes.

\begin{figure}
\centering
  \epsfig{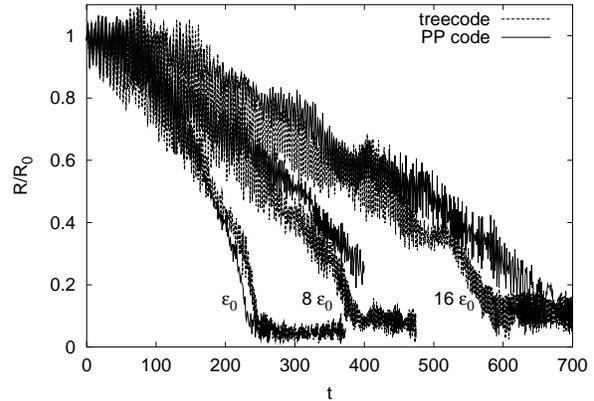}
\caption{\label{codecompfig} Comparison of results from the PP code with results from the treecode, at different values of $\eps$. For all cases shown here is \mbox{$N = 80\,000$} and \mbox{$M_{BH} = 0.000528$}. The PP simulation with \mbox{$\eps = 8\eps_0$} has been already shown in Fig.~\ref{obs_fig}.}
\end{figure}

In Fig.~\ref{codecompfig} selected treecode runs with \mbox{$N = 80\,000$} and increasing $\eps$ are compared with the direct code runs having the same values of $N$ and $\eps$. At the same time, the figure shows how the infall time increases (and implicitly how $\lnL$ decreases), as $\eps$ increases. 
Fig.~\ref{codecompfig} and Table~\ref{loltab} show that the results from the treecode, the PP code, and the PM code are in good agreement. The agreement of the results from the three methods, and the scaling of $\lnL$ with $\eps$, will be further studied quantitatively in Section~\ref{lnLsec}.

\begin{figure}
\centering
  \epsfig{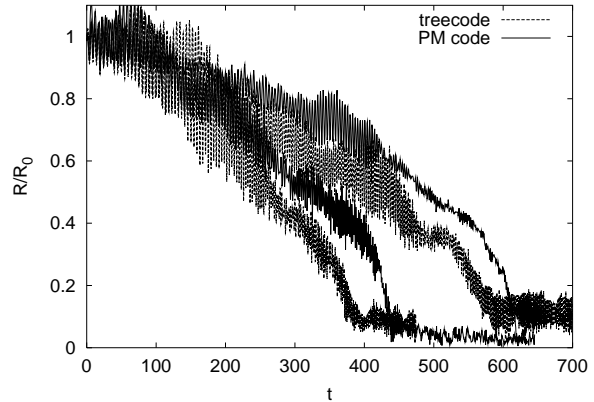}
\caption{\label{codecomp1fig} Comparison of PM results with treecode results. PM simulations have cell size $l$ equal to resp. $10\eps_0$ and $23\eps_0$; softening parameters in the treecode runs are resp. $8\eps_0$ and $16\eps_0$. In all the above cases, is \mbox{$N = 80\,000$} and \mbox{$M_{BH} = 0.000528$}.}
\end{figure}

In order to understand how the cell length $l$ of the PM code and the softening parameter $\eps$ of the PP code and treecode relate with each other, we compare in Fig.~\ref{codecomp1fig} the results from the PM code and treecode simulations with $80\,000$ particles. The BH infall as shown in Fig.~\ref{codecomp1fig} depends on the value of $l$ or $\eps$. Remarkably, $l$ and $\eps$ seem to play the same role not only qualitatively, but also quantitatively: in a PM run, a given value of $l$ induces an infall which is very similar to the infall, in a treecode run, with $\eps$ assuming that same value. In Section~\ref{lnLsec} this relation will be studied further.

\subsection{The effect of softening/grid} \label{Sect:softening}

\begin{table}
\caption{$\lnL$ versus $\eps$ from PP code, treecode, and PM code runs. For 
the PP code and treecode runs is \mbox{$N = 80\,000$}. For the PM code runs is 
\mbox{$N = 2$\,million}.The reference value for the accuracy parameter is 
\mbox{$\eps_0 = 0.003735$}.}
\label{loltab}
  \begin{center}
    \begin{tabular}{ccc}
     $\eps$/$\eps_0$ & PP code & treecode  \\
      \hline \hline
     $0$             & 6.6     &           \\
     $0.01$          & 6.0     &           \\
     $0.1$           & 5.3     & 5.7       \\
     $1$             & 4.8     & 4.7       \\
     $2$             & 3.5     & 4.1       \\
     $8$             & 2.8     & 3.0       \\
     $16$            & 1.8     & 2.0       \\
     $32$            &         & 1.6       \\
    \end{tabular}
     \begin{tabular}{cc}
      $l$/$\eps_0$ & PM code  \\
      \hline \hline
                   &          \\
                   &          \\
                   &          \\
           2.5     &    4.1   \\
           5       &    3.8   \\
           10      &    3.0   \\
           23      &    2.2   \\
    \end{tabular}
    \end{center}
\end{table}

The influence of the softening parameter on the BH dynamics has been studied by
running a number of simulations with the three codes. In Table~\ref{loltab} we 
report the value of $\lnL$ obtained from our simulations. For the PP code
and treecode simulations, we increase $\eps$ from 0 to \mbox{$32 \eps_0 = 
0.1195 \simeq 0.96$\,pc}. For the PM code, we increase $l$ from 2.5\,$\eps_0$ 
to 23\,$\eps_0$. In all cases is \mbox{$M_{BH} = 0.528 \cdot 10^{-3} \simeq 
62\,300$\,M$_\odot$}. 

For the PP code and the treecode, we selected \mbox{$N = 80\,000$} as a 
suitable value. The relaxation time is for this value of $N$ \mbox{$t_x \simeq 
0.1 N/ \ln N \cdot R/v_{typ} \simeq 2000$}, about one order of magnitude larger
than the typical BH infall time, so that the system is collisionless, and we 
can confidently use the treecode to simulate it. With this choice for $N$, the 
BH mass is \mbox{$M_{BH}/m \simeq 42.3$}, (see Table~\ref{allrunTREEtab}). As a
cross-check, we ran two PP runs with \mbox{$N = 16\,000$} which, as expected, 
gave incorrect results (see Table~\ref{allrunPPtab}). This is due to both a too
small $M_{BH}/m$ ratio, and a too short relaxation time (\mbox{$t_x \simeq 
400$} in this case). We did not increase $\eps$ above $\simeq 0.12$, since at 
this point $\eps$ is already much bigger than $b_{min}$ (see 
Table~\ref{allrunTREEtab}), and the infall time is now close to $t_x$.

For the PM code simulations, we used \mbox{$N = 2$\,million} in order to have enough particles to fill all the cells, even for the simulations with a small $l$. As Table~\ref{allrunPMtab} shows, for \mbox{$l = 0.076 \rmn{pc} \simeq 2.5 \eps_0$} the average number of particles per cell is already \mbox{$N_c = 0.1$}. Since a PM simulation gives incorrect results for \mbox{$N_c \ll 1$} (see also the discussion at the end of Section~\ref{Nscalesec}), we did not decrease $l$ below 2.5\,$\eps_0$.

The decrease of the value of $\lnL$ as $\eps$ or $l$ increases is clear from Table~\ref{loltab}. In the next sub-section we focus on the relation between $\Lambda$ and $\eps$, and provide a fitting formula for $\lnL(\eps)$. We use hereafter $\eps$ to refer either to the softening length of the PP and treecode, or the cell size of the PM code. As shown on Fig.~\ref{codecomp1fig} and discussed above, these two parameters play the same role even quantitatively in affecting $\lnL$. In this respect, we refer to $\eps$ as a generic accuracy parameter.

\subsection{Determination of $\bmath{\lnL}$} \label{lnLsec}
We will study here the relation between $\eps$ and $\lnL$. As just said above, in this context $\eps$ will be used as the accuracy parameter, and it will refer to either the softening length used in the PP and treecode, or to the cell size in the PM code.

A mathematical expression for the relation between $\eps$ and $\lnL$ can be found by considering how softening affects two body scattering. The role of $\eps$ is to prevent too close stellar encounters. In this respect, the effect of introducing a softening length is to increase the minimal impact parameter. Hence, we can define an effective impact parameter \mbox{$b_{eff} = b_{min} + \eps$}, and we modify equation~(\ref{lolgeneq}) to become:
\begin{equation}
   \lnL = \ln \frac{b_{max}}{b_{eff}} = \ln \frac{b_{max}}{b_{min} +\eps} \, .
\label{thfiteq} \end{equation}
We will now fit this equation with the values reported in Table~\ref{loltab}. In order to perform the fit, we change equation~(\ref{thfiteq}) in a more suitable form, as follows:
\begin{equation}
    \lnL = \ln b_{max} -\ln (b_{min} +\eps) = K - \ln( a + \eps) \, .
\label{fiteq} \end{equation}
We will refer to $b_{max}$ and $b_{min}$ as the theoretical values of the maximal and minimal impact parameters, as they can be obtained from equations~(\ref{loleq}) and~(\ref{lolgeneq}), and $K$ and $a$ as the corresponding experimental values obtained with the fit.

\begin{figure}
\begin{center}
  \epsfig{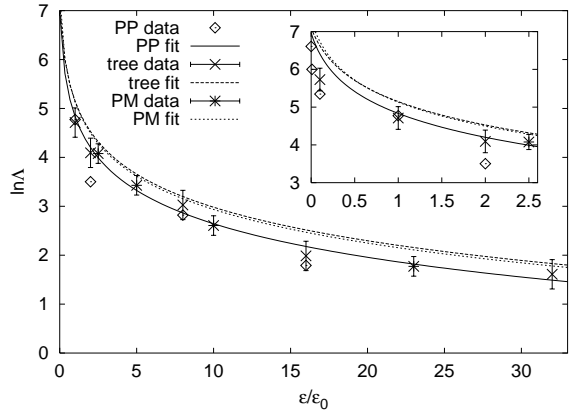}
  \caption{\label{fitfig} $\lnL$ vs $\eps$, and best fit for \mbox{$ \lnL = K - \ln( a + \eps)$}. Values for $K$ and $a$ are given in Table~\ref{fittab}. The inset in the figure is a magnification of the low $\eps$ region. In all cases is \mbox{$M_{BH} = 0.000528$}. For the PP and treecode runs is \mbox{$N = 80\,000$}, for the PM code runs is \mbox{$N = 2\,000\,000$}. Error bars are omitted from the PP values to improve readability. For the same reason, $\lnL$ values for \mbox{$\eps/\eps_0 < 1$} are shown only in the inset.}
\end{center}
\end{figure}

The best fits for $K$ and $a$ with respect to simulation values are reported in Table~\ref{fittab} for all codes. Such fits have been performed with a fixed value for $R_0$, i.e.\ the \mbox{$R_0 = R_{hm}$}. In fact, the not perfectly circular orbit of the BH results in an oscillatory behaviour for the BH galactocentric radius. In this case, having $R_0$ fixed could not be an appropriate choice for the fit. We checked whether having $R_0$ as a free parameter in the fit leads to different results in $\lnL$. We obtained values for $\lnL$ within the error bars in Fig.~\ref{fitfig}, and values for $R_0$ within \mbox{$R_{hm} \pm 0.05$}. We can conclude that, although the galactocentric BH radius does not decreases smoothly, but in an oscillatory fashion, having $R_0$ fixed to the actual initial BH radius in the simulations leads to correct fits for the value of $\lnL$. With respect to the PM values, a further peculiarity is that when the BH enters the finest grid area, i.e.\ approximately at \mbox{$R = 0.3$}, the value of $l$ decreases (see Section~\ref{sec:superbox} and Fig.~\ref{SBgridfig}). This causes $b_{eff}$ to become smaller, increasing the value of $\lnL$. In fact, a fit of the PM data limited to values of \mbox{$R > 0.3$} gives values of $\lnL$ systematically higher by \mbox{$\simeq 0.3 \simeq 2 \Delta (\lnL)$}. 

\begin{table}
\caption{ Best values for the parameters $K$ and $a$, and error on $\lnL$ for the fit of \mbox{$ \lnL = K - \ln( a + \eps)$}.}
\label{fittab}
\begin{center}
\begin{tabular}{crrr}
                  &  PP code        &  treecode          & PM code          \\ 
\hline \hline
$K$               & -0.94 $\pm$ 0.21 & -0.64 $\pm$  0.10 & -0.59 $\pm$ 0.05 \\
$a \cdot 10^{-3}$ &  0.80 $\pm$ 0.28 &  0.88 $\pm$  0.20 &  0.74 $\pm$ 0.08 \\
$\Delta (\lnL)$   &             0.6  &              0.3  &             0.2  \\  
\end{tabular}
\end{center}
\end{table}

From the PP code value of $K$ in Table~\ref{fittab} we obtain for $b_{max}$ the experimental value \mbox{$b_{max}^E = e^K \simeq 0.39$}. This value is quite smaller than what one would expect. Since $b_{max}$ has the meaning of the maximal impact parameter, a natural choice is to assign it a value of the order of the system size, which in our case would result in \mbox{$b_{max} = 2$}. The maximal radius for dynamical friction in our system is then about one quarter of what it is customarily assumed. PM and treecode values are slightly higher, but still quite smaller than \mbox{$b_{max} = 2$}. Also $a$ is smaller than the theoretical value \mbox{$b_{min} = G \cdot (M_{BH} + m)/v_{typ}^2 = 1.41 \cdot 10^{-3}$}, by a factor 3. The $a$ value for all codes is perfectly consistent.

An explanation for the discrepancy between the values of $b_{max}$ and $b_{max}^E$ is that the BH, while moving to the Galactic centre, is off-centre with respect to the density peak (in fact the BH is spiralising towards it). With respect to the BH position, the density distribution is then asymmetric. This density peak has clearly a greater influence on the BH dynamics, contributing more than the other regions of the system to the dynamical friction on the BH. This leads to a value of $b_{max}$ affected by the galactocentric BH radius. This approach is studied in detail by \citet{HasFun02}, who propose the galactocentric radius as a value for $b_{max}$ in the context of the spiralisation of satellite galaxies. 

In our simulations, the galactocentric radius varies from \mbox{$R \simeq 0.9$} at the beginning of a simulation, to \mbox{$R \simeq 0$} at the end of it. The value of $b_{max}^E$ that we find is within this range, and it can be interpreted as an order 0 estimate of a maximal impact parameter that depends on the galactocentric BH radius.

\begin{figure}
\begin{center}
  \epsfig{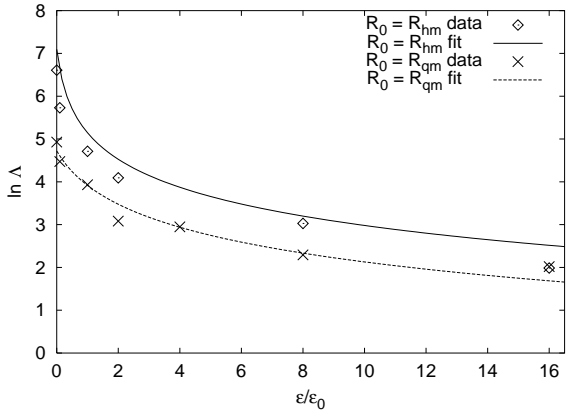}
  \caption{\label{qmasslolfig} Comparison of $\lnL$ vs $\eps$ at different initial galactocentric BH radii. The smaller values of $\lnL$ for the cases with \mbox{$R_0 = R_{qm}$} indicate that $b_{max}$ is influenced by the galactocentric BH radius.}
\end{center}
\end{figure}

In order to explore this aspect further, we simulated the infall of the same BH, starting at the quarter mass radius \mbox{$R_{qm} \simeq  0.43$}, for $\eps$ ranging from 0 to \mbox{$16 \eps_0$}. What we expect is a smaller value of $b_{max}^E$, hence smaller values of $\lnL$. All simulations are performed with the treecode, except for the \mbox{$\eps = 0$} case, which is simulated with the PP code. Our results are in Fig.~\ref{qmasslolfig}.
We can see there how smaller are the values of $\lnL$ for the cases when the BH starts at the quarter mass radius. A fit on these data gives \mbox{$K \simeq -1.1$}, which implies \mbox{$b_{max}^E \simeq 0.33$}, which is smaller than the value of $b_{max}^E$ obtained for the BH starting from the half mass radius. Our findings support the argument of \citet{HasFun02}.

\subsection{Varying black hole mass}\label{Sect:mass}

\begin{figure}
\begin{center}
  \epsfig{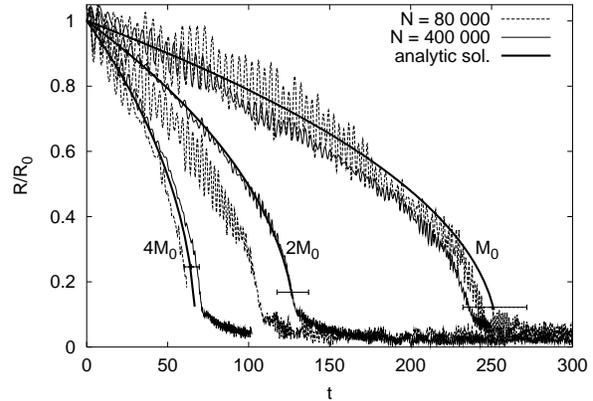}
  \caption{\label{varmfig} Black hole infall for different values of the BH mass, and different values of $N$. Simulations are performed with the treecode. Simulation results are compared with the analytic solution, equation~(\ref{Rvsteq}), with $\lnL$ obtained from equation~(\ref{fiteq}) and Table~\ref{fittab}. The error bars at the bottom of the analytic curves correspond to a variance \mbox{$\Delta(\lnL) = \pm 0.3$}.}
\end{center}
\end{figure}

We also studied the effect of a variable BH mass on the value of $\lnL$. We simulated, using the treecode, the infall of a BH of mass
two times and four times larger than the default mass \mbox{$M_0 = 0.528
\cdot 10^{-3} \simeq 0.62 \cdot 10^5 {\rm M}_\odot$}. We studied this
infall in both the 80\,000 particles configuration, and the 400\,000
particles configuration. In all cases, we used our standard value for
$\eps$, i.e. \mbox{$\eps_0 = 0.003735$}. In Fig.~\ref{varmfig} the
distance $r$ of the BH from the centre of mass of the system is shown
for all the cases mentioned above, together with the $M_{BH} = M_0$
cases. From equation~(\ref{fiteq}) and Table~\ref{fittab}, the
appropriate value for $\lnL$ in the above cases is:
\begin{eqnarray*}
    \lnL & = & K - \ln(a + \eps_0) \pm \Delta  = \\
         &   & -0.64 - \ln(0.00088 + 0.003735) \pm 0.3 \simeq 4.7 \pm 0.3\ .
\end{eqnarray*}
We also show in Fig.~\ref{varmfig} the analytic curve $r(t)$, as given by equation~(\ref{Rvsteq}), with $\lnL = 4.7$. An error bar gives, for each analytical curve, the spread corresponding to a variance \mbox{$\Delta(\lnL) =  0.3$}. 

The results shown in Fig.~\ref{varmfig} are consistent with the
hypothesis that a variation in the BH mass has a little effect in the
value of $\lnL$. In fact, $\lnL$ shows a logarithmic dependence on $M_{BH}$ through the parameter $b_{min}$, which depends linearly on $M_{BH}$ (see equations~(\ref{loleq}) and~(\ref{lolgeneq})). Assuming that also the experimental value $a$ depends linearly on $M_{BH}$, we obtain \mbox{$\lnL \simeq 4.6 \pm 0.3$}, and \mbox{$\lnL \simeq 4.4 \pm 0.3$} respectively for the 2~$M_{BH}$ case and the 4~$M_{BH}$ case. This results in a small displacement towards the right of the corresponding analytic curves in Fig.~\ref{varmfig}, which does not
affect the conclusions that can be drawn from the figure. The
theoretical curve fits very well with the $M = 2\,M_0$, $N = 400\,000$
case. The other simulation curves are within, or very close to, the
error in $r(t)$ associated to the error in $\lnL$. We can
conclude that a variation in the mass of the infalling object has little influence in the value of $\lnL$, which is important in view of extending this
work to the case of the infall of a star cluster.

The fitting formula for \mbox{$\lnL$ vs $\eps$} was obtained from
simulations with \mbox{$M_{BH} = M_0$}. This formula
predicts $\lnL$ for the cases with \mbox{$M_{BH} > M_0$} with a very good accuracy, showing that it can be applied in a more general context, in order to forecast the value of the Coulomb logarithm. 

Fig.~\ref{varmfig} also shows a damping in the BH infall at very small values of $R$, especially for the \mbox{$N = 400\,000$} case. This effect, described in Section~\ref{softsec}, is clearer in the \mbox{$N = 400\,000$} case, since the particle density is higher in this case, compared to the \mbox{$N = 80\,000$} case.

\subsection{Comparison with related work} \label{relworksec}

\citet{MilMer01} study the dynamical evolution of two black holes, each one at the centre of a power law cusped galaxy core. They simulate the merging of the two galaxies, which results in the two black holes forming a hard binary at the centre of the merged galaxy. In Section~3 of their paper they discuss the value of $\lnL$ in their simulations. They measure the decay rate of the two black holes, and compare this value with theoretical estimates. When they compare their experimental decay rate with an estimate for the case of the infall of an isolated black hole, they find a theoretical estimate about 6 times lower than the measured value, under the assumption that \mbox{$\lnL \simeq 1.6$}. If the value of $\lnL$ is not theoretically pre-determined, and is instead obtained from the decay rate equation, the result is \mbox{$\lnL \simeq 10$}. Similarly, they compare the experimental value with an estimate for the case of two mutually spiralling spherical distributions of matter. In this case they assume \mbox{$\lnL \simeq 1.0$}, and obtain an estimate for the decay rate about a factor of 2 lower than the observed value. Determining $\lnL$ from the measurement would give in this case \mbox{$\lnL \simeq 1.87$}. The values of $\lnL$ that we find are between the two values above.

The value for $\lnL \simeq 1$ that they assume in their theoretical estimates, comes from a derivation that they present in appendix~A of the same work. This derivation is based on results of \citet{Mao93}. Under the assumption that the stellar density obeys a power law centered on the BH position:
\begin{equation}
    \rho(r) = \rho_0 \left( \frac{r}{b_{min}} \right)^{-\alpha} \ ,
\label{MMrhoeq} \end{equation}
they obtain \mbox{$\Lambda \simeq 1/\alpha \simeq 1$}, which actually implies \mbox{$b_{max} \simeq b_{min}$}, whereas it is customary to consider \mbox{$b_{max} \gg b_{min}$}. 

Their assumption in fact is valid only when the BH is close to the centre of the power law distribution. In their context this is true when: 1) the separation between the two BHs is much larger than the half mass radius of the two galaxies. In this case each BH is at the centre of its own galaxy, and at the same time its motion is not yet heavily perturbed by the other galaxy. 2) the BH binary has hardened, and occupies the centre of the merged galaxy. 

During the transient phase, when the two BHs have not yet formed a binary, the density distribution that affects the motion of the BHs is double-cusped, with a BH in each of the two cusps. This is substantially different from the density distribution modelled by equation~(\ref{MMrhoeq}).

This qualitative argument would make the density distribution in equation~(\ref{MMrhoeq}) inapplicable during the transient phase, and could explain why \citet{MilMer01} find a higher than expected value of $\lnL$ in the transient. The analytical evaluation of $\lnL$ according to the technique used by them is by no means trivial, when symmetry arguments cannot be straightforwardly applied. We will address this issue in future developments of the present work; the theory of linear response of \citet{ColPal98} could be very useful in this context. 

\section{Applications to star clusters}\label{Sect:discussion}

Recent observations of the Galactic Centre have revealed a population
of very young clusters with ages less than 10 Myr. The presence of
such stars inside the inner pc of the Galaxy is puzzling, as the
strong tidal field in the Galactic centre easily prevents star
formation. The origin of these stars is therefore vividly debated
(\citealt{Ger01}; \citealt{McMPor02}; \citealt{KimMor02}). \citet{Mor93} proposed that a star cluster at some distance from the Galactic centre could spiral-in due to dynamical friction (see also \citet{Ger01}). The efficiency of dynamical friction depends sensitively on the actual value of the Coulomb logarithm $\lnL$. 

\subsection{Sinking of massive black holes in the Galactic centre}

We performed $N$-body simulations for a large range of
conditions. In Section~\ref{Nscalesec} we varied the number of particles, in Section~\ref{Sect:softening} we varied the size of the object, and in Section~\ref{Sect:mass} we varied its mass. With direct $N$-body simulations we measured the actual value of the Coulomb logarithm $\lnL$. We study the behaviour of $\lnL$ for various types of $N$-body solvers and particle numbers. We also study the behaviour of $\lnL$ as a function of the softening length $\eps$. Only the direct $N$-body code can perform a true measurement of the Coulomb logarithm, because it is able to resolve even the smallest length scales and time scales. This, however, makes the direct code very slow and, even using the very fast GRAPE-6 special purpose device, we are able to perform simulations with only $10^5$ particles. This is a small number compared to the actual number of stars in the Galactic centre. With approximate methods (treecode and particle-mesh) we are able to increase the number of particles up to 2~million. The cost of this is a lower accuracy in calculating stellar motion below a typical length scale $\eps$. We studied how this length scale influences $\lnL$, by affecting the value of the minimal impact parameter.

\subsection{Young dense clusters in the Galactic centre}
\label{sec:ycgc}

The study of the dependence of $\lnL$ on $\eps$ described above is also of astronomical interest, because $\eps$ can be interpreted as the typical length of a finite size infalling object. Based on this, our analysis of the dependence of $\lnL$ on $\eps$ can be seen as a first approach to the study of the infall of a star cluster of typical size $\eps$ toward the Galactic centre. We found (see Fig.~\ref{fitfig}) that the value of $\lnL$ decreases quite rapidly as $\eps$ increases, with the logarithm argument \mbox{$\Lambda \propto 1/\eps$}. The typical size of the compact young clusters observed in the Galactic bulge is \mbox{$\simeq 0.3$}\,pc~\citep{FigKim99}, which corresponds to \mbox{$\eps \simeq 10 \eps_0$}. With this value of $\eps$, from equation~(\ref{fiteq}) and Table~\ref{fittab}, we obtain \mbox{$\lnL \simeq 2.9$}, about 60\% less than the value for a point mass. The infall time is roughly doubled. For our choice of object mass, \mbox{$M \simeq$ 62~300 M$_\odot$}, and initial galactocentric radius, \mbox{$R_0 \simeq 7$ pc}, we have an infall time that increases from \mbox{$\simeq 6$\,Myr} for the point mass, to \mbox{$\simeq 12.5$\,Myr} for an object of typical size \mbox{$\simeq 10 \eps_0 \simeq 0.3$\,pc}. 

We also studied the uncertainty associated with the maximal impact parameter $b_{max}$. We found that for an infall to the Galactic centre, the infalling object is mostly influenced by the density peak at the Galactic centre itself. A good choice for $b_{max}$ is then \mbox{$b_{max} \simeq \beta R_0$}, where $R_0$ is the initial galactocentric radius, and \mbox{$\beta \simeq 0.5$}

\section{Conclusions}

We simulated the evolution of a massive particle in a sea of lighter
particles in a self gravitating system.  The main goal of this work is
to obtain an accurate value of the Coulomb logarithm ($\lnL$).
This helps us to understand the dynamics of the Galactic bulge and the
rate at which intermediate mass black holes sink to the Galactic
centre. We also study the effect of the finite size of the inspiraling object.  

We ran both $N$-body particle-particle (PP) simulations, softened treecode simulations, and particle-mesh (PM) simulations. The comparative simulations are performed for 80\,000 particles, and all result in the same value of $\lnL$.  For a point particle near the Galactic centre we find \mbox{$\lnL = 6.6 \pm 0.6$}. In addition we measure the change in the Coulomb logarithm with respect to the softening parameter $\eps$, which reveals \mbox{$\Lambda \propto
1/\eps$}. We argue that $\eps$ can be interpreted as the typical length of a finite size object, such as a star cluster, so that $\lnL$ as a function of $\eps$ can be seen as a first approximation of the dependence of the Coulomb logarithm on the size of an infalling star cluster.

We also observed a value of the maximal impact parameter $b_{max}$ different from the customarily assumed value, which is proportional to the system size. We found that our results are more consistent to a value of $b_{max}$ linearly dependent on the BH galactocentric radius, which is in agreement with \citet{HasFun02}.

We performed simulations with up to 2~million particles using a treecode. The obtained value of $\lnL$ does not depend on the
number of particles. Apparently, 80\,000 particles is already enough
to eliminate any granularity for our choice of initial conditions.
The results of the treecode, at the low $N$-limit, are in excellent
agreement with the PP simulations, and we find the same scaling with
respect to $\eps$. Increasing the black hole mass reduces the time
scale for spiral-in as was expected from theory (see \citealt{McMPor02}).

Finally we compared the results of our PP and treecode simulations
with a particle mesh (PM) method. We compared the methods for $N$ up to 2~million. The results of our PP, treecode, and PM calculations are in good agreement.  The cell size in the PM model is directly comparable to the softening length $\eps$ in the PP and tree methods.

\section*{acknowledgements}
  
We thank Douglas Heggie, Piet Hut, Jun Makino, and Steve McMillan for stimulating discussions. We acknowledge KNAW under grant 95-BTN-15, NWO under Spinoza grant 08-0, NOVA under grant V-37, DFG-grant FE564/1-1, EC through grant HPRI-1999-CT-00026. SPZ is a KNAW fellow.

%\bibliographystyle{mn2e}
%\bibliography{../../ilmio}
%

\end{document}